\documentclass[superscriptaddress,aps,prl,showkeys,showpacs,twocolumn,10pt]{revtex4}
\usepackage[OT1,T1]{fontenc}
\usepackage{graphicx}
\usepackage{rotating}
\begin{document}

\title{Branching Ratios for the Decay of $d^*(2380)$}
\date{\today}

\newcommand*{\Edin}{School of Physics and Astronomy, University of Edinburgh,
James Clerk Maxwell Building, Peter Guthrie Tait Road, Edinburgh, EH9 3FD, UK}

\newcommand*{\PITue}{Physikalisches Institut, Eberhard--Karls--Universit\"at
  T\"ubingen,  
Auf der Morgenstelle~14, 
D-72076 T\"ubingen, Germany}

\newcommand*{\Kepler}{Kepler Center for Astro and Particle Physics, University
of T\"ubingen, Germany}
\newcommand*{\Tomsk} {Department of Physics, Tomsk State University, 36 Lenina
  Avenue, Tomsk, 634050, Russia}

\author{M.~Bashkanov}\affiliation{\Edin}\affiliation{\PITue}\affiliation{\Kepler}
\author{H.~Clement}     \affiliation{\PITue}\affiliation{\Kepler}
\author{T.~Skorodko}    \affiliation{\Tomsk}

\begin{abstract}
Based on measurements, the branching ratios for the decay of the recently
discovered dibaryon resonance $d^*(2380)$ into two-pion production channels
and into the $np$ channel are evaluated. Possibilities for a decay into the
isoscalar single-pion channel are discussed. Finally, the electromagnetic
decay of $d^*(2380)$ is considered.  
\end{abstract}


\maketitle

\section{Introduction}
 
Recent WASA experiments at CELSIUS \cite{prl2009} and COSY
\cite{prl2011,isofus} on the basic double-pionic 
fusion to deuterium identified a narrow isoscalar resonance structure with
mass  $ m \approx$ 2.37 GeV and width $\Gamma \approx$ 70 MeV in the
total cross section of the reactions $pn \to d\pi^0\pi^0$ and $pn \to
d\pi^+\pi^-$. The differential distributions are consistent
with a spin-parity assignment of $J^P =  3^+$ to this structure. Subsequent
measurements of two-pion production reactions, where the two colliding
nucleons do not fuse to deuterium, but stay unbound, also show this resonance
effect, if the reaction contains isoscalar parts \cite{pp0-,np00,np+-}.

The final proof for this structure to represent a genuine $s$-channel
resonance has been achieved by polarized $\vec{n}p$ scattering in the energy
region of interest. The obtained analyzing power data produce a pole in the
coupled $^3D_3 - ^3G_3$ partial waves at $(2380\pm10) - i(40\pm5)$ MeV, if
included in the SAID data base with subsequent partial-wave analysis
\cite{prl2014,npfull,RW}. Henceforth, this state has been denoted by $d^*(2380)$
following the convention used for nucleon excitations.

The golden reaction channel for the observation of $d^*(2380)$ turned out to be
$pn \to d\pi^0\pi^0$, since there the background from conventional processes
due to $t$-channel Roper and $\Delta\Delta$ excitations is smallest. Since
WASA has been the only detector with a nearly full solid angle coverage for
both charged and neutral particles, which was placed at a hadron accelerator,
it is of no surprise that there were no data for this channel from previous
measurements. Thus it was left to the WASA collaboration to reveal the
pronounced Lorentzian energy dependence sitting upon an only small background
in the total cross section of this channel.

The experimental Dalitz plot at resonance in this golden channel $pn \to
d\pi^0\pi^0$ points to a
$\Delta\Delta$ excitation in the intermediate state \cite{prl2011}, which
means that $d^*(2380)$ behaves asymptotically like a bound $\Delta\Delta$
system. We note that already in 1964 Dyson and Xuong \cite{dyson} predicted
exactly such a state, based on SU(6) symmetry considereations, with a value
for the mass remarkably close to the one of the now observed dibaryon
resonance. 
Later-on Kamae and Fujita \cite{kamae} as well as Goldman {\it et al.}
\cite{goldman} predicted such a state, though the latter with a markedly lower
mass. Only recent modern quark-model calculations see this state properly near
the experimental mass \cite{ping,shen,zhang1,zhang2,chen}. Also new
relativistic Faddeev-type calculations carried out by Gal and Garcilazo by
use of hadronic interactions find this state at the correct mass \cite{GG1,GG2}.

If we account just for the well-known momentum dependence of the width of the
$\Delta(1232)$ resonance, then we expect for a conventional $\Delta\Delta$
system bound by 80 MeV a decay width of about 160 MeV \cite{BBC,dong}. This is
more than twice that observed for $d^*(2380)$. Hence it is of no surprise that
theoretical calculations predict a too large width for this state. Until very
recently Gal and Garcilazo \cite{GG2} came closest with about 100 MeV for the
width, if they allow for all decay channels discussed in the next section.
This discrepancy in the width might indicate some exotic contribution, which
hinders the decay of $d^*(2380)$ -- such as hidden color as discussed in
Refs. \cite{zhang1,BBC}. In fact, very recently it has been shown in
Ref. \cite{dong,zhang3} that the experimentally observed small width can be
reconciled theoretically, if the hidden color aspect is taken into account.  

To clarify the experimental situation, we examine in the
following, whether all major hadronic decays of $d^*(2380)$ have been
identified and understood or whether a substantial decay branch has escaped
detection so far.

\section {Decay Channels and Widths}

\subsection {Hadronic Decays}
We consider the following reaction scenario as suggested by the data on
two-pion production \cite{prl2011,isofus,pp0-,np00,np+-}:

\begin{equation}
pn \to d^*(2380) \to \Delta\Delta \to (NN\pi\pi)_{I=0},
\end{equation}

where $d^*(2380)$ denotes an $s$-channel resonance both in $pn$ and
$\Delta\Delta$ systems. By this scenario, we neglect a possible direct decay
$d^*(2380) \to NN\pi$, but we shall come back to this point in the next to next
section. Note that an intermediate $N\Delta$ configuration is excluded by
isospin.  

\subsubsection{$NN\pi\pi$ and $np$ channels}

First, we consider the possible decay channels in the
scenario of eq. (1). In particular we estimate the partial decay width into
the elastic $pn$ channel. 

The cross section of the  isoscalar two-body resonance process $pn \to
d^*(2380) \to \Delta\Delta$ is given by the relativistic Breit-Wigner formula
\cite{PDG}: 
\begin{eqnarray}
\sigma_{pn \to \Delta\Delta} = \frac {4 \pi} {k_i^2}  \frac {2J+1} {(2s_p +
  1) (2s_n + 1)}  
\frac {m_{d^*}^2 \Gamma_i
\Gamma_f} {(s-m_{d^*}^2)^2 +m_{d^*}^2 \Gamma^2},
\end{eqnarray}
where $k_i$ denotes the initial center-of-mass momentum.

As best estimates for mass and width of the resonance, we take the average over
the results from elastic scattering and two-pion production, {\it i.e.}
$m_{d^*}$ = 2.375 GeV and $\Gamma$ = 75 MeV.

With $J$ = 3 and $s_p = s_n$ = 1/2, the peak cross section at $\sqrt s =
m_{d^*}$ = 2.375 GeV ($k_i$ = 0.73 GeV/{\it c}) is then 
\begin{equation}
\sigma_ {pn \to \Delta\Delta}(peak) = \sigma_0 \frac  {\Gamma_i \Gamma_f }
{\Gamma^2}
\end{equation}
with 
\begin{equation}
\sigma_0 = 16.1~{\rm mb ~~(unitarity~limit)}.
\end{equation}

Since we also have

\begin{equation}
\Gamma = \Gamma_i + \Gamma_f,
\end{equation}
we get from (3) and (5):
\begin{equation}
\Gamma_i = \Gamma \left( \frac 1 2 \pm \sqrt { \frac 1 4 - \frac
  {\sigma_{pn \to \Delta\Delta}(peak)} {\sigma_0 }}\right).
\end{equation}

In general, partial and total widths are momentum dependent
quantities. Therefore, branching ratios, as we will derive below for d*(2380),
	have been defined by convention to be quoted at the Breit-Wigner
	resonance mass or at the resonance pole, see Particle Data Group
	recommendations \cite{PDG}. In case of d*(2380) both coincide within 
	uncertainties. Hence, we need to consider the resonance cross sections
	only at its peak position. Also, in all relevant WASA experiments the
        energy resolution has been similar, in the range of 10 - 20 MeV, which
	is small compared to the resonance width. Hence the smearing effect on
	the resonance curve is small and in particular comparable in all
	measured channels, so that this effect tends to cancel in the
        branching ratios to large extent.

To estimate $\sigma_{pn \to \Delta\Delta}(peak)$ consider the total cross
sections of all channels, where the isoscalar $\Delta\Delta$ system can decay
into:  
\begin {itemize}
\item (i) $d\pi^0\pi^0$ and $d\pi^+\pi^-$: \\
  Due to isospin rules we expect
\begin{equation}
\sigma_{d\pi^+\pi^-}(d^*)~=~2~\sigma_{d\pi^0\pi^0}(d^*). 
\end{equation}
However, due to the isospin violation in the pion mass, the available phase
space is somewhat smaller for charged pion production than for the production
of the lighter neutral pions. In Ref. \cite{isofus} it has been shown that
this results in a resonance cross section, which is lower by about 20$\%$ in
case of the $d\pi^+\pi^-$ channel. Hence we have
\begin{eqnarray}
\sigma_a := \sigma_{d\pi^+\pi^-}(d^*) + \sigma_{d\pi^0\pi^0}(d^*) \\\nonumber
\approx 2.6~\sigma_{d\pi^0\pi^0}(d^*).
\end{eqnarray}

The peak cross section of the $pn \to d\pi^0\pi^0$ reaction at  $\sqrt s$ = 2.37
GeV  has been measured to be 0.27 mb \cite{isofus}. This includes the
contributions of the 
$t$-channel $\Delta\Delta$ and Roper excitations. Accounting for these
background effects, the pure resonance cross section in this channel amounts to
about 0.24~mb,  
{\it i.e.}, $\sigma_a \approx$ 0.62~mb.
Since the cross sections of the three fusion reactions $pn \to d\pi^0\pi^0$,
$pn \to d\pi^+\pi^-$ and $pp \to d\pi^+\pi^0$ measured in Ref.~\cite{isofus}
are closely connected by isospin, only a small uncertainty of about 10$\%$ in
absolute normalization has been quoted. This leads to
\begin{equation}
\sigma_a \approx 0.62(6)~{\rm mb}.
\end{equation}

\item (ii) $pp\pi^0\pi^-$, $nn\pi^+\pi^0$, $np\pi^0\pi^0$ and $np\pi^+\pi^-$
  --- isoscalar parts: \\

First, we consider the $pp\pi^0\pi^-$ channel. Though both the $pp$ pair and the
$\pi^0\pi^-$ pair are isovector pairs, together they may couple to total
isospin $I = 0$. Hence the isoscalar resonance $d^*(2380)$ may also
decay into the isoscalar part of the $pp\pi^0\pi^-$ channel. In fact, 
the decay of the resonance into the $pp\pi^0\pi^-$ channel proceeds via the
same intermediate $\Delta^+\Delta^0$ system as the $d\pi^0\pi^0$ channel
does. From isospin coupling we expect that the resonance decay into the
$pp\pi^0\pi^-$ system should be half of that into the $np\pi^0\pi^0$ system,
which is in the order of 0.2 mb -- see next to next paragraph. In
fact, a recent measurement \cite{pp0-} of this channel by WASA-at-COSY is in
agreement with a resonance contribution of 0.10(1) mb in the total cross section
at $\sqrt s$~=~2.37~GeV.

The $nn\pi^+\pi^0$ channel is just the isospin mirrored one to the
$pp\pi^0\pi^-$ channel. Hence it has to have the same resonance contribution.   
 
In a recent paper \cite{WF}, F\"aldt and Wilkin present an estimate of the
resonance cross section in the $pn \to pn\pi^0\pi^0$ reaction. According to
their calculation based on final state interaction theory, the expected peak
cross section in the deuteron breakup channel $pn\pi^0\pi^0$ is about 85$\%$
that of the non-breakup channel $d\pi^0\pi^0$, {\it i.e.} about 0.2 mb. 
Recently also Albaladejo and Oset \cite{oset} estimated the expected resonance
cross sections  in $pn \to pn\pi^0\pi^0$ and $pn \to pn\pi^+\pi^-$ using a more
elaborate theoretical procedure. Their result for the $pn \to pn\pi^0\pi^0$
channel is compatible with that from Ref.~\cite{WF}. In fact, a recent
measurement at WASA \cite{np00} shows that the data are in accordance with a
contribution of $d^*(2380)$ with a strength of 0.20(3)~mb.

The resonance effect in the isoscalar part of the $np\pi^+\pi^-$ channel is
composed of the configurations, where either both $np$ and $\pi^+\pi^-$ pairs
couple each to $I~=~0$ or both pairs each to $I~=~1$. The first case gives
just twice the contribution in the $np\pi^0\pi^0$ channel. The latter case
provides the same situation as in the $pp\pi^0\pi^-$ channel. Hence, we have 

\begin{eqnarray}
\sigma_{np\pi^+\pi^-}(d^*)\approx 2 \sigma_{np\pi^0\pi^0}(d^*) +
\sigma_{pp\pi^0\pi^-}(d^*) \\ 
\approx 0.50(8)~{\rm mb}. \nonumber
\end{eqnarray}

Note that in these non-fusion channels there is no ABC effect, {\it i.e.} no
low-mass enhancement in the $\pi\pi$-invariant mass spectra
\cite{pp0-,np00,np+-}. Hence the 
phase-space reduction due to different masses of charged and neutral pions
as discussed above for the $d\pi^+\pi^-$ channel does not play a significant
role here.
 
Our estimate for the resonant $pn \to pn\pi^+\pi^-$ cross section
is in good agreement with that of Ref. \cite{oset}, where, however, only the
contributions with $I~=~0$ coupled nucleon and pion pairs were considered. Our
result also is compatible with available measurements for this channel, see
Ref.~\cite{np+-}, including the newest results from HADES \cite{hades}.

In total we have from these four reactions
\begin{eqnarray}
\sigma_b := \sigma_{np\pi^+\pi^-}(d^*) + \sigma_{np\pi^0\pi^0}(d^*)\\ \nonumber
+ \sigma_{pp\pi^0\pi^-}(d^*) + \sigma_{nn\pi^+\pi^0}(d^*) \\ \nonumber
= 0.50(8)~{\rm mb} + 0.20(3)~{\rm mb}\\\nonumber
+ 0.10(1)~{\rm mb} + 0.10(1)~{\rm mb}  \\\nonumber
= 0.90(13)~{\rm mb}. \nonumber
\end{eqnarray}
\end {itemize}

Altogether we get as an estimate 
\begin{equation}
\sigma_{pn \to \Delta\Delta}(peak) = \sigma_a + \sigma_b  \approx 1.5(2)~{\rm mb}.
\end{equation}
Putting this into eq. (6) and selecting the minus sign in front of the root (see
discussion below) we obtain
\begin{equation}
\Gamma_i = 8(1) {\rm MeV  ~~~~~~~~for}~~\Gamma = 75~{\rm MeV},
\end{equation}
which in turn corresponds to a resonance cross section in the elastic $pn$
channel of only
\begin{equation}
\sigma_{pn \to pn} \approx 0.17(2)~{\rm mb},
\end{equation}
if we apply eq. (3) for the incident channel.

From the SAID partial-wave analysis of elastic $np$ scattering including the
new WASA data on polarized $\vec{n}p$ scattering in the energy region of
$d^*(2380)$ an elastic partial width of $\Gamma_i$ = 10(2) MeV has been derived
corresponding to a branching ratio of 12(3)$\%$ \cite{npfull}. This result
agrees reasonably well with the value obtained above in eq.~(13). 

We note in passing that the other solution of eq. (6) -- the one with the
 plus sign -- leads to the complementary result, namely $\Gamma_i$ = 67 MeV --
 thus implying that the resonance would be predominantly elastic, {\it i.e.},
 mainly decay into the elastic channel and only weakly into the
pion-production channels. This solution is at obvious variance with elastic
$np$ scattering data.

From the peak cross sections given under (i) and (ii) as well as from eqs. (3)
- (12) we may readily calculate the branching ratios BR := $\Gamma_j /
\Gamma$ for the decay of the resonance into the individual $NN\pi\pi$
channels. The results are listed in Table 1.

\begin{table}
\caption{Experimental branching ratios (BR) of the $d^*$ resonance into its
  decay channels based on eqs. (1 - 3) and (12) and the peak cross sections
  given 
  under  (i) and (ii).}  
\begin{tabular}{lllll}
\\ 
\hline


 &decay channnel&BR&derived from&\\ 

\hline

& $np$ & 12(3) $\%$ & measurement \cite{npfull} &  \\
& $d\pi^0\pi^0$ & 14(1) $\%$& measurement \cite{isofus} &  \\
& $d\pi^+\pi^-$ & 23(2) $\%$ & measurement \cite{isofus} &  \\
& $np\pi^+\pi^-$ & 30(5) $\%$ & measurement \cite{np+-,hades} &  \\ 
& $np\pi^0\pi^0$ & 12(2) $\%$ & measurement \cite{np00} &  \\ 
& $pp\pi^0\pi^-$ & ~~6(1) $\%$ & measurement \cite{pp0-} &  \\ 
& $nn\pi^+\pi^0$ & ~~6(1) $\%$ & isospin symmetry &  \\
\hline
 \end{tabular}\\
\end{table}

\begin{figure} 
\centering
\includegraphics[width=0.9\columnwidth]{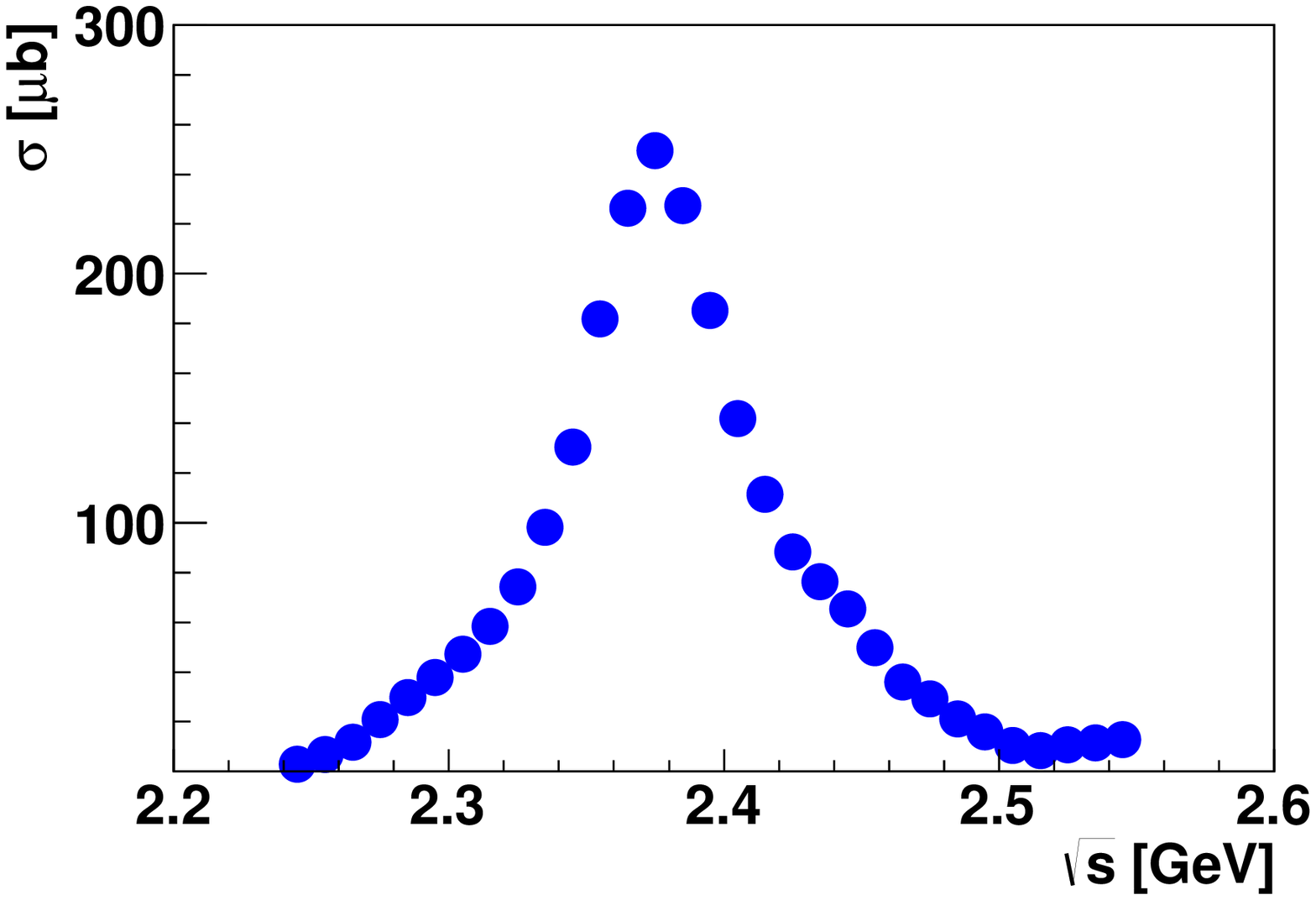}
\includegraphics[width=0.9\columnwidth]{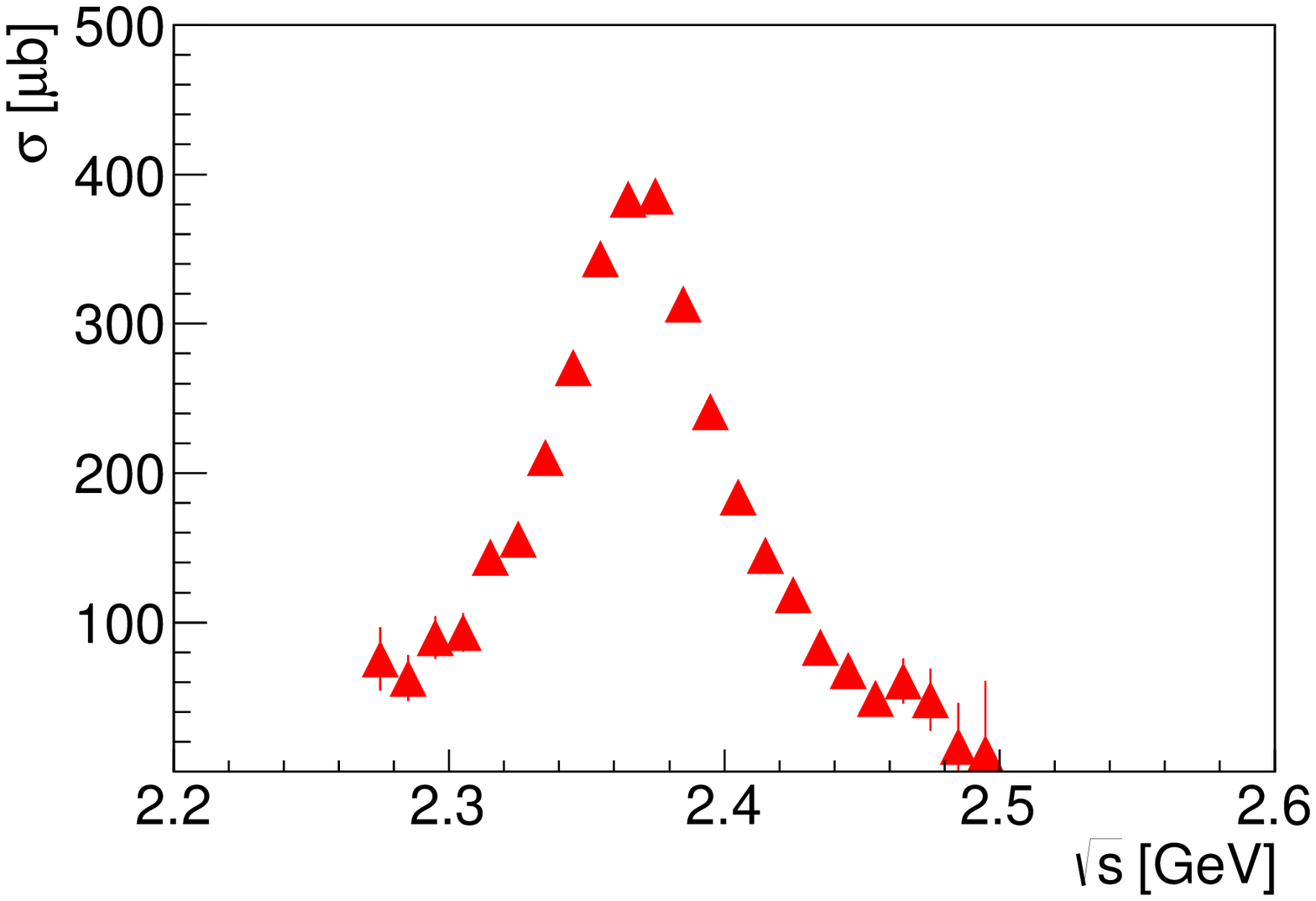}
\includegraphics[width=0.9\columnwidth]{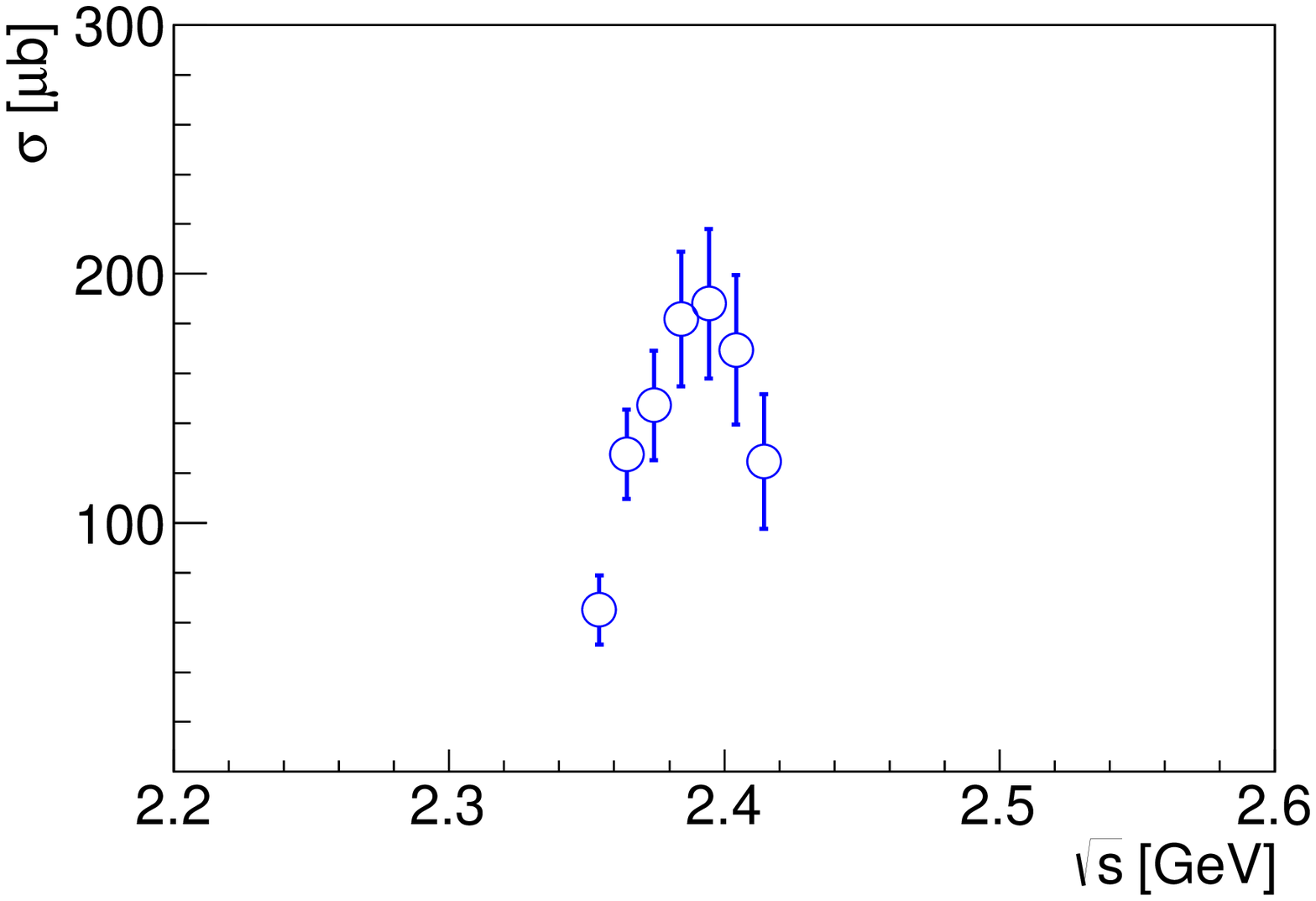}
\includegraphics[width=0.9\columnwidth]{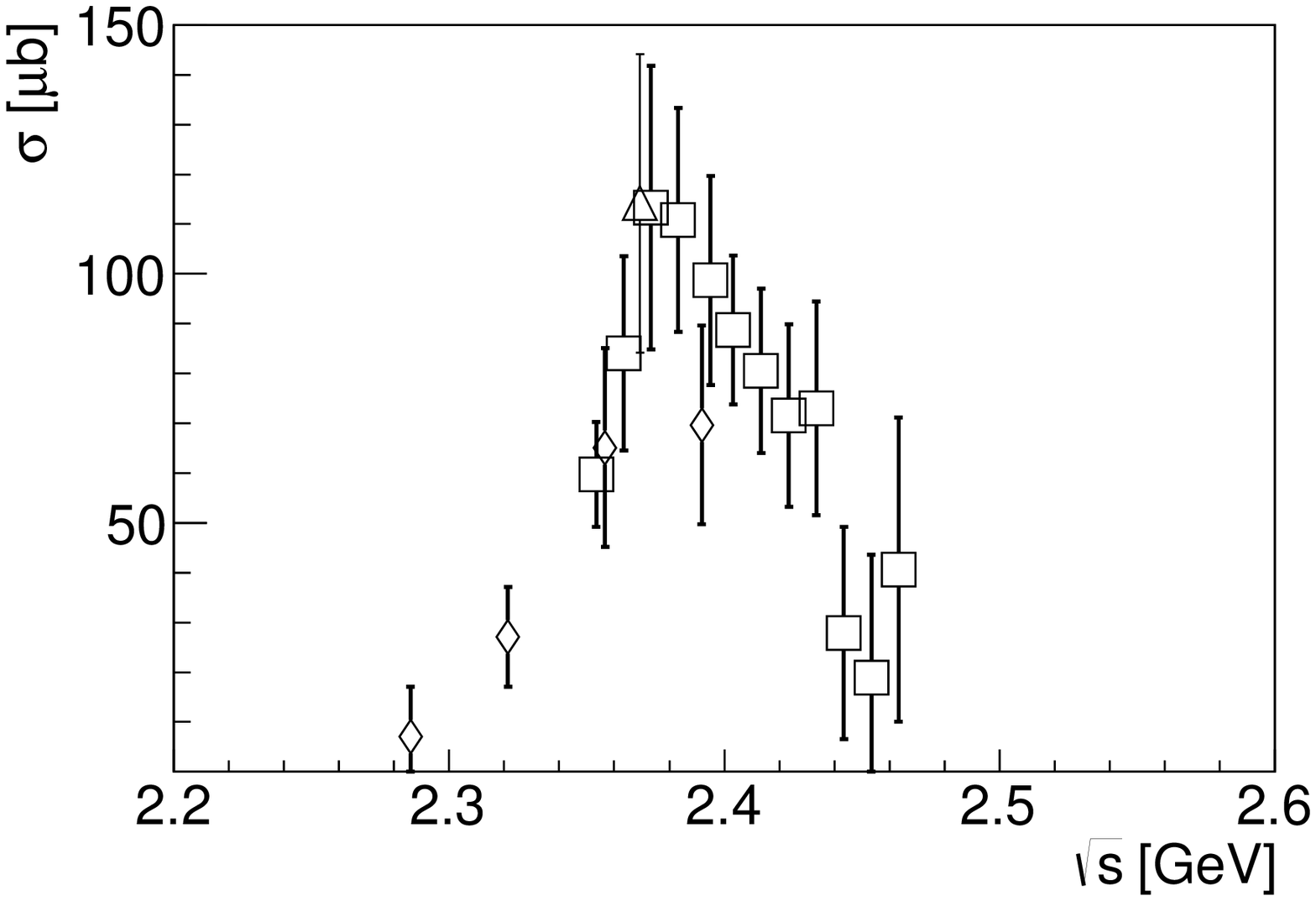}
\caption{\small Background subtracted cross sections for the reactions $pn
  \to d\pi^0\pi^0$, $pn \to d\pi^+\pi^-$ (isoscalar part), $pn \to
  pn\pi^0\pi^0$ and $pn \to 
  pp\pi^0\pi^-$ (from top to bottom) as obtained from WASA measurements
  \cite{prl2011,isofus,np00,pp0-} and Refs. \cite{KEK,RAL} by subtraction of the
  conventional background as given in Refs. \cite{prl2011,isofus,np00,pp0-}. 
}
\label{fig1}
\end{figure}

The experimental branching ratios listed in Table 1 add up to 103(15)$\%$, which is consistent within the assumption in eq. (1) that the $d^*$
      decay proceeds solely via the intermediate $\Delta\Delta$ system and the
      $np$ channel. From the data on differential observables, in particular
      Dalitz plots and $N\pi$-invariant mass spectra, we know that these are
      the dominant decay channels. But this does not exclude that there might
      be other small decay channels. {\it E.g.}, it has been proposed that
      there might be a 5$\%$ decay branch $d^* \to d \sigma$ with an unusual
      light and narrow $\sigma$ particle, in order to describe the low-mass
      enhancement (ABC effect), see Refs. \cite{Kuk,natureABC}.
  
      From the fact that we have two independent results for the elastic
      partial width, namely the one obtained from the SAID partial-wave
      analysis and the one derived in eq. (8), we may obtain an upper limit
      for decay branchings not covered by eq. (1). If we account for such
      branching by a partial width $\Gamma_s$, then $\Gamma$ in eq.(6) is to be
      replaced by $(\Gamma - \Gamma_s)$. Assuming then, e.g., $\Gamma_s/\Gamma$
      to be in the order of 10 $\%$ leads subsequently to a reduction of our
      result for the elastic partial width from 8 to 7 MeV, i.e. the agreement
      with the SAID results fades away. Thus we estimate that the sum of
      branchings missed by eq. (1) can not be larger than 10 $\%$.

As an alternative to the consideration of peak cross sections we may also
exam the full resonance effect in the various reaction channels. Since
the signal-to-background ratio is largest at the resonance peak, an integration
over the full resonance region is naturally more sensitive to the assumed
background and also to the negligence of interferences of the resonance with
the background. However, statistics improves that way and this method is largely
independent of the experimental energy resolution. Hence this method may serve
as a cross-check of the results obtained by the peak cross section
method. Fig.~1 shows the background subtracted cross sections for the
reactions $pn \to d\pi^0\pi^0$, $pn \to d\pi^+\pi^-$ (isoscalar part), $pn \to
pn\pi^0\pi^0$ and $pn \to pp\pi^0\pi^-$ (from top to bottom) as obtained from
WASA measurements \cite{prl2011,isofus,np00,pp0-} and Refs. \cite{KEK,RAL} by
subtraction of the conventional background given in
Refs. \cite{prl2011,isofus,np00,pp0-}. In order to keep the background problem
as small as possible, but to account still sufficiently for any energy
smearing due to the experimental resolution, we do not integrate over the full
resonance region, but only 
from $m_{d^*} - \Gamma/2$ to $m_{d^*}+ \Gamma/2$. If we normalize then to the
branching ratio of 14 $\%$ for the $d^*$ decay into $d\pi^0\pi^0$ we obtain
21(1) $\%$, 11(1) $\%$ and 7(1)$\%$ for the branching ratios the $d^*$ decay
into $d\pi^+\pi^-$, $pn\pi^0\pi^0$ and $pp\pi^0\pi^-$ channels,
respectively. These results agree very well with those from the peak cross
section method given in Table~1.

\subsubsection{$NN\pi$ channel}

One such channel, which has not yet been investigated experimentally, is
$d^*(2380) 
\to NN\pi$, {\it i.e.} concerns isoscalar single-pion production. Since
single-pion production in $NN$ collisions is either purely isovector or
isospin mixed, the isoscalar part has to be obtained by combination of various
cross section measurements. Most favorable appears the combination
\cite{isosarantsev}: 
\begin{eqnarray}
\sigma_{NN \to NN\pi}(I = 0) = 3 ( 2 \sigma_{np \to pp \pi^-} - \sigma_{pp \to
  pp\pi^0} ),
\end{eqnarray}
which derives from isospin decomposition of single-pion production \cite{Bys}.
Experimentally the most difficult part is the measurement of the $np \to
pp\pi^-$ reaction, since it affords either neutron beam or target. Technically
this may be achieved by use of deuteron beam or target and measurement of the
above reaction in the quasi-free mode. Since this necessitates, however,
exclusive and kinematically complete measurements, in order to obtain reliable
results, the data base on that is sparse, see Ref. \cite{isosarantsev}. In
particular, there are no data in the region of the $d^*(2380)$ resonance. 

Also theoretically, it is difficult to construct a process of sizable cross
section, where 
$d^*(2380)$ can decay into the single-pion channel. As mentioned at the
beginning of this section, an intermediate $\Delta N$ system is isospin
forbidden. Hence the next simple candidate configuration would be $N^*(1440)
N$. However, spin-parity $J^P = 3^+$ of  $d^*(2380)$ would require a $d$-wave
between $N^*(1440)$ and $N$. Since the resonance energy is just at the
$N^*(1440) N$ threshold, the probability for such a decay must be tiny already
from the kinematical point of view \cite{natureABC}. Another possibility would
be, if $d^*$ decays via the resonance structure $D_{12}(2150)$ with $I(J^P) =
1(2^+)$ at the $N\Delta$ threshold, as proposed in Refs. \cite{GG1,GG2,Kuk},
which in turn can decay into the $NN$ system. This would give then the
scenario $d^* \to D_{12}(2150) \pi \to NN\pi$.
       Also, the following process $np \to d^* \to np \to NN\pi$ appears to be
       possible in principle. However, aside from the fact this affords another
       reaction step beyond elastic scattering via the $d^*$ resonance, this
       process has to proceed 
       with the $^3D_3$ partial wave, which is not dominant in single pion
       production. 

Whatsoever, a careful experimental investigation of this issue appears to be
appropriate. Since WASA at COSY has finished its experimental program, dedicated
measurements on that issue are no longer possible. However, the existing data
base at WASA taken for various purposes contains data also on the desired
single-pion production channels. A corresponding data analysis is in progress.

\subsection{Electromagnetic Decays}

An electromagnetic excitation of the deuteron groundstate to the $d^*(2380)$
resonance is highly informative, since its transition formfactor gives
access to size and structure of this resonance.

Judging just from the electromagnetic coupling constant, we expect
electromagnetic decays to be suppressed already by two order of magnitudes --
as is borne out, {\it e.g.} in the decay of the $\Delta$ resonance. A
technical feasible excitation of $d^*(2380)$ would start by photo or electro
excitation from the deuteron groundstate. A real or virtual photon would need
then to transfer two units of angular momentum, {\it i.e.} be of E2 or C2
multipolarity, which lowers the transition probability further. In addition, the
overlap in the wavefunctions of $d$ and $d^*(2380)$ enters profoundly. We are
aware of two 
theoretical calculations dealing with such a scenario \cite{wong,qing}, where
cross sections in the range pb/sr - nb/sr are predicted for the forward
angular range. These are two orders lower than conventional processes and
hence it appears very difficult to sense the resonance excitation under usual
conditions. In this respect the reaction $\gamma d \to d\pi^0\pi^0$ appears
       to be attractive, since the conventional processes there are expected
       to be particularly small \cite{Fix}.

A way out could be polarization measurements. The situation looks similar to
the one in elastic $np$ scattering. As we have shown above, the $d^*(2380)$
resonance contribution is about 0.17 mb, which is more than two orders below
the total elastic cross section. However, with help of the analyzing power,
which consists only of interference terms in partial waves, it was possible to
filter out reliably the resonance contribution. 

The analogous case in electro or photo excitation of $d^*(2380)$ constitute 
measurements of the polarization of the outgoing proton in the reactions
$\gamma d \to n\vec{p}$ and $\gamma^* d \to n\vec{p}$, respectively, where
$\gamma^*$ stands for a virtual photon created in inelastic electron
scattering on the deuteron. As in the analyzing power of $np$ scattering the
angular dependence of the resonance effect in the polarization of the outgoing
proton should be proportional to the associated Legendre polynomial
$P_3^1(cos~\Theta)$ \cite{npfull}. Therefore, the maximal resonance effect is
expected to be at a scattering angle of $\Theta$ = 90$^\circ$.

In fact, such an effect has already been looked for previously by Kamae
{\it et al.} in corresponding data from the Tokyo electron synchrotron
\cite{kamae1,kamae2,ikeda1,ikeda2}. In order to describe the observed large
polarizations in 
the region of $d^*(2380)$ they fitted a number of resonances to the data,
among others also a $J^P = 3^+$ state. However, presumably due to the limited
data base they only obtained very large widths for these resonances in the
order of 200 - 300 MeV -- as one would expect from conventional $\Delta\Delta$
excitations. 

Recently, new polarization measurements from JLAB appeared \cite{JLAB}. Their
lowest energy point is just in the $d^*(2380)$ region and is compatible
with a maximal polarization of $P = -1$. It confirms thus the old Tokyo results
in the sense that in this region there is a build-up of a very large
polarization, which rapidly decreases both towards lower and higher energies,
see Fig.~1 in Ref.~\cite{JLAB}. Of course, a dedicated measurement over the
region of interest is needed, in order to see, whether a narrow structure with
the width of $d^*(2380)$ can be observed in this observable.

\section{Conclusions}

We have considered all $NN\pi\pi$ channels, into which the isoscalar dibaryon
resonance $d^*(2380)$ can decay. For all of these channels there exist
meanwhile experimental data, which show the $d^*(2380)$ resonance contribution
and thus deliver the corresponding decay branchings. These branchings are
compatible with what one expects from isospin coupling, if the intermediate
state is a $\Delta\Delta$ configuration. This in turn agrees with the result
from the Dalitz plot in the golden channel $d\pi^0\pi^0$, where the background
situation is optimal. We add that there is no sign of this
resonance observed in isovector $NN\pi\pi$ channels \cite{isofus,iso,deldel}.

So the only possible hadronic decay channel, which missed so far a careful
inspection, is the $NN\pi$ channel -- though we know of no simple mechanism,
by which 
$d^*(2380)$ could decay into such an isoscalar configuration. However, since
such a scenario has not yet been examined experimentally, a dedicated
experimental investigation appears to be in order.

The electromagnetic decay is expected to be tiny compared to the hadronic
decay branchings. As we also pointed out, the $d^*(2380)$ contribution in
deuteron disintegration processes will be even small compared to the background
from conventional processes. A way out will possibly be the measurement of
polarization observables. In particular, the polarization of the outgoing
proton or neutron appears to be very promising as discussed above. Forthcoming
measurements at MAMI could possibly give a decisive answer on that.

\section{Acknowledgments}

We acknowledge valuable discussions on this matter with Stanley J. Brodsky,
A. Gal, J. Haidenbauer, C. Hanhart, F. Hinterberger, T. Kamae, E. Oset,
I. Strakovsky, G.J. Wagner, C. Wilkin, A. Wirzba, R. Workman and Z. Zhang.   
This work has been supported by the Forschungszentrum J\"ulich (COSY-FFE), 
DFG (CL 214/3-1) and STFC.


\begin{thebibliography}{9}
 
\bibitem{prl2009} M. Bashkanov {\it et al.}, Phys. Rev. Lett. {\bf 102}, 052301
  (2009).  
\bibitem{prl2011} P. Adlarson {\it et al.}, Phys. Rev. Lett. {\bf 106},242302
  (2011). 
\bibitem{isofus} P. Adlarson {\it et al.}, Phys. Lett. B {\bf 721},229 (2013).
\bibitem{pp0-} P. Adlarson {\it et al.}, Phys. Rev. C {\bf 88}, 055208 (2013).
\bibitem{np00} P. Adlarson {\it et al.}, Phys. Lett: B {\bf 743}, 325 (2015). 
\bibitem{np+-} H. Clement, M. Baskanov and T. Skorodko, Proc. STORI 2014,
  Phys. Scr., in press; arxiv:1506.00557 [nucl-ex].
\bibitem{prl2014} P. Adlarson {\it et al.}, Phys. Rev. Lett. {\bf 112}, 202301
  (2014). 
\bibitem{npfull} P. Adlarson {\it et al.}, Phys. Rev. C {\bf 90}, 035204 (2014).
\bibitem{RW} R. Workman, EPJ Web Conf. {\bf 81}, 02023 (2014).
\bibitem{dyson} F.J. Dyson and N.-H. Xuong, Phys. Rev. Lett. {\bf 13}, 815
  (1964); ibid. {\bf 14}, 339 (1965) (errata).   
\bibitem{kamae} T. Kamae and T. Fujita, Phys. Rev. Lett {\bf 38}, 471 (1977).
\bibitem{goldman} T. Goldman {\it et al.}, Phys. Rev. C {\bf 39}, 1889 (1989).
\bibitem{ping} H. Huang, J. Ping and F. Wang, Phys. Rev. C {\bf 89}, 034001
  (2014) and references therein. 
\bibitem{shen} Q. B. Li and P. N. Shen, J. Phys. G {\bf 26}, 1207 (2000).
\bibitem{zhang1} F. Huang, Z. Y. Zhang, P. N. Shen and W. L. Wang,
  arxiv:1408.0458[nucl-th].
\bibitem{zhang2} X. Q. Yuan, Z. Y. Zhang, Y. W. Yu, P. N. Shen, Phys. Rev. C
  {\bf 60},045203 (1999).
\bibitem{chen} Hua-Xing Chen {\it et al.}, arxiv:1410.0394 [hep-ph].
\bibitem{GG1} A. Gal and H. Garcilazo, Phys. Rev. Lett. {\bf 111}, 172301
  (2013).   
\bibitem{GG2} A. Gal and H. Garcilazo, Nucl. Phys. A {\bf 928}, 73 (2014).
\bibitem{BBC} M. Bashkanov, Stanley J.  Brodsky and H. Clement,
Phys. Lett. B {\bf 727},438 (2013).
\bibitem{dong} Yubing Dong, Pengnian Shen, Fei Huang and Zongye Zhang,
  arxiv:1503.02356 [nucl-th].
\bibitem{zhang3} F.Huang, P. N. Shen, Y.- B. Dong and Z. Y. Zhang,
  arxiv:1505.05395 [nucl-th].  
\bibitem{PDG} K.A. Olive {\it et al.} (Particle Data Group), Chin. Phys. C
  {\bf 38}, 090001 (2014).
\bibitem{WF} G. F\"aldt and C. Wilkin, Phys. Lett. B {\bf 701}, 619 (2011).
\bibitem{oset} M. Albaladejo and E. Oset, Phys. Rev. C {\bf 88}, 014006 (2013).
\bibitem{hades} G. Agakishiev {\it et al.}, arxiv:1503.04013 [nucl-ex].
\bibitem{Kuk} M. Platonova and V. Kukulin, Phys. Rev. C {\bf87}, 025202 (2013). 
\bibitem{natureABC} M. Bashkanov, H. Clement and T. Skorodko, arxiv:1502.07500
  [nucl-ex].
\bibitem{KEK} T. Tsuboyama  {\it et al.}, Phys. Rev. C {\bf 62}, 034001
  (2000).
\bibitem{RAL} D. C. Brunt, M. J. Calyton and B. A. Westwood, Phys. Rev. {\bf
    187}, 1856 (1969).
\bibitem{isosarantsev} V. V. Sarantsev {\it et al.}, Eur. Phys. J. A {\bf 43},
  11 (2010). 
\bibitem{Bys} J. Bystricky  {\it et al.}, J. Phys. {\bf 48}, 1901 (1987).
\bibitem{wong} Chun Wa Wong, Phys. Rev. C {\bf 61}, 064011 (2000).
\bibitem{qing} Di Qing, He-Ming Sun and Fan Wang, Chin. Phys. Lett. {\bf 18},
  885 (2001).
\bibitem{Fix} M. Egorov and A. Fix, Nucl. Phys. A {\bf 933}, 104 (2015).
\bibitem{kamae1} T. Kamae {\it et al.}, Phys. Rev. Lett. {\bf 38}, 468 (1977).
\bibitem{kamae2} T. Kamae {\it et al.}, Nucl. Phys. B {\bf 139}, 394 (1978).
\bibitem{ikeda1} H. Ikeda {\it et al.}, Phys. Rev. Lett. {\bf 42},1321 (1979).
\bibitem{ikeda2} H. Ikeda {\it et al.}, Nucl. Phys. B {\bf 172}, 509 (1980).
\bibitem{JLAB} K. Wijesooriya {\it et al.}, Phys. Rev. Lett. {\bf 86}, 2975
  (2001). 
\bibitem{iso} T. Skorodko {\it et al.}, Phys. Lett. B {\bf 679}, 30 (2009).
\bibitem{deldel} T. Skorodko {\it et al.}, Phys. Lett. B {\bf 695}, 115 (2011).
\end{thebibliography}
\end{document}